\documentclass[reprint,
twocolumn,
superscriptaddress,
twocolappendix,
]{aastex702}

\usepackage{times}
\usepackage{mathtools}
\usepackage{graphicx}
\usepackage{dcolumn}
\usepackage{bm}
\usepackage{soul}
\usepackage{verbatim}
\usepackage{cprotect}
\usepackage{url}
\usepackage{subfigure}
\usepackage{tikz}
\usepackage{afterpage}
\usepackage{nicefrac}
\usepackage{tabularx}
\usepackage{multirow}
\usepackage[english]{babel}
\usepackage{natbib}
\usepackage{hyperref}
\usepackage{comment}
\usepackage{float}
\usepackage{placeins}
\usepackage{ulem}
\usepackage{booktabs}
\graphicspath{{figures/}}
\usepackage[acronym]{glossaries}
\newacronym{GW}{GW}{Gravitational wave}
\newacronym{SNR}{SNR}{signal-to-noise ratio}
\newacronym{BBH}{BBH}{binary black hole}
\newacronym{LVK}{LVK}{LIGO-Virgo-KAGRA}
\newacronym{ROC}{ROC}{Receiver operating characteristic}
\newacronym{AUC}{AUC}{area under the ROC curve}

\usepackage{xcolor}

\newcommand{\gstlal}{GstLAL~}

\begin{document}

\title{Turning Selection Bias into a New Statistic: Asymmetry and Fringes of Lensed Gravitational Waves}
\author{Juno C. L. \surname{Chan}}
\affiliation{Center of Gravity, Niels Bohr Institute, Blegdamsvej 17, 2100 Copenhagen, Denmark}
\affiliation{Department of Physics, The Chinese University of Hong Kong, Shatin, New Territories, Hong Kong}
\email[show]{chun.lung.chan@nbi.ku.dk}

\author{Elwin K.~Y. \surname{Li}}
\affiliation{Department of Physics, The Chinese University of Hong Kong, Shatin, New Territories, Hong Kong}
\email{elwinlikayau@link.cuhk.edu.hk}

\date{\today}

\begin{abstract}
\glspl{GW} propagating across the Universe can be distorted by intervening lenses, producing characteristic wave-optics signatures.
These effects are not incorporated into current matched-filtering searches and can therefore substantially reduce detection efficiency.
The underlying mechanism is the signal-consistency test, which measures deviations between the observed and expected \gls{SNR} distributions.
We report a systematic study of the impact of gravitational lensing on signal-consistency tests, using a two-image interference model and representative signal-consistency statistics in both the time and frequency domains.
We find that overlapping images produce sharply localized \gls{SNR} residuals in the time domain, breaking the reflection symmetry of the autocorrelation about the merger time.
Although interference fringes produce broadband \gls{SNR} residuals in the frequency domain, both domains exhibit a minimum for high-mass signals when the fringe structure becomes comparable to the signal bandwidth.
This highlights a limitation of current strategies that search for distorted lensed signals only among \gls{GW} events detected by matched-filtering searches.
Exploiting this asymmetry, we develop a statistic that separates selected down-ranked lensed injections from noise-background triggers and paired nonlensed injections, without extending the template bank.
Our results establish the first connection between lensing observables and detection, paving the way for the discovery of lensed \glspl{GW}.
\end{abstract}

\section{Introduction}\label{Section: Introduction}

\glspl{GW} propagating through the Universe can be gravitationally lensed by intervening mass distributions~\citep{Nakamura:1997sw,Takahashi:2003ix}.
In the geometric-optics regime, strong lensing produces multiple images with different magnifications, arrival times, and Morse phases~\citep{Wang:1996as,Dai:2017huk,Ezquiaga:2020gdt}.
For \glspl{GW} observed by ground-based detectors, lenses with redshifted masses of roughly $10$--$10^5\,M_{\odot}$ can produce observable wave-optics effects, with diffraction and interference causing frequency-dependent distortions in the waveform amplitude and phase~\citep{Jung:2017flg}.
Diffraction becomes important when the wavelength is comparable to or longer than the characteristic gravitational length scale of the lens, while images that overlap in time interfere to produce frequency-dependent distortion in the waveform~\citep{Takahashi:2003ix,Liu:2023emk,Lo:2024wqm,Ezquiaga:2025gkd,Liu:2023ikc}.
These wave-optics signatures provide probes of compact objects and dark-matter structure across a range of scales~\citep{Fairbairn:2022xln,Tambalo:2022wlm,Savastano:2023spl,Cheung:2024ugg,Harikumar:2026pnu,Vujeva:2026xfh,Zumalacarregui:2026uqs}.
They can also break lensing degeneracies relevant to cosmography and probe modifications to \gls{GW} propagation beyond general relativity~\citep{Cremonese:2021puh,Chen:2024xal,Liu:2024xxn,Menadeo:2024uoq,Garoffolo:2026hyx}.

Searches for distorted images in the first four \gls{LVK} observing runs rely on \glspl{GW} detected by matched-filtering searches~\citep{Hannuksela:2019kle,LVKlensing:2023bwz,Janquart:2023mvf,Janquart:2024ztv,Chakraborty:2024mbr,LVKlensing:2025cwb,Chan:2025kyu,Chakraborty:2025maj,Goyal:2025eqo,Hu:2025lhv,Chakraborty:2025pxt,Cheung:2026pky}.
The impact of lensing distortions on matched-filter searches is often approximated from the \gls{SNR} alone~\citep{Mishra:2023ddt,Ubach:2025dob,Barsode:2026bcs,Gholap:2026fjy,Shan:2026beq}.
Matched-filtering searches, however, also use signal-consistency tests to assess whether the matched-filter response agrees with that expected from the template and to suppress noise transients that can produce large \glspl{SNR}~\citep{Allen:2004gu,Allen:2005fk,Venumadhav:2019tad,Aubin:2020goo,Chu:2020pjv,DalCanton:2020vpm,Cannon:2021a,Allene:2025saz,Huang:2025pkd}.
Together with the recovered \gls{SNR}, these tests contribute to the ranking statistics used to distinguish astrophysical signals from the noise background.
Lensing-induced waveform distortions can therefore reduce search sensitivity through both a loss of \gls{SNR} and an increase in signal inconsistency, causing otherwise significant triggers to be down-ranked~\citep{Chan:2024qmb}.
However, the mapping between lensing observables and these consistency penalties remains poorly understood, limiting our ability to predict lensing-induced losses in search sensitivity and the associated selection effects.

In this \emph{Letter}, we establish a connection between lensing observables and time- and frequency-domain signal-consistency tests widely adopted in matched-filtering searches~\citep{Allen:2004gu,Messick:2017a,Chu:2020pjv,Allene:2025saz}.
We show that the residual morphology differs qualitatively between the two domains and identify the regime in which lensed signals have a reduced consistency penalty.
We then exploit the localized and asymmetric time-domain residual to construct an asymmetry statistic that separates lensed signals from the noise background.
These results motivate a practical route to mitigating lensing-induced selection effects in current matched-filtering searches without extending the template bank.

\section{Two-image Interference Model}

\begin{figure*}[t!]
    \centering
    \includegraphics[width=0.9\textwidth]{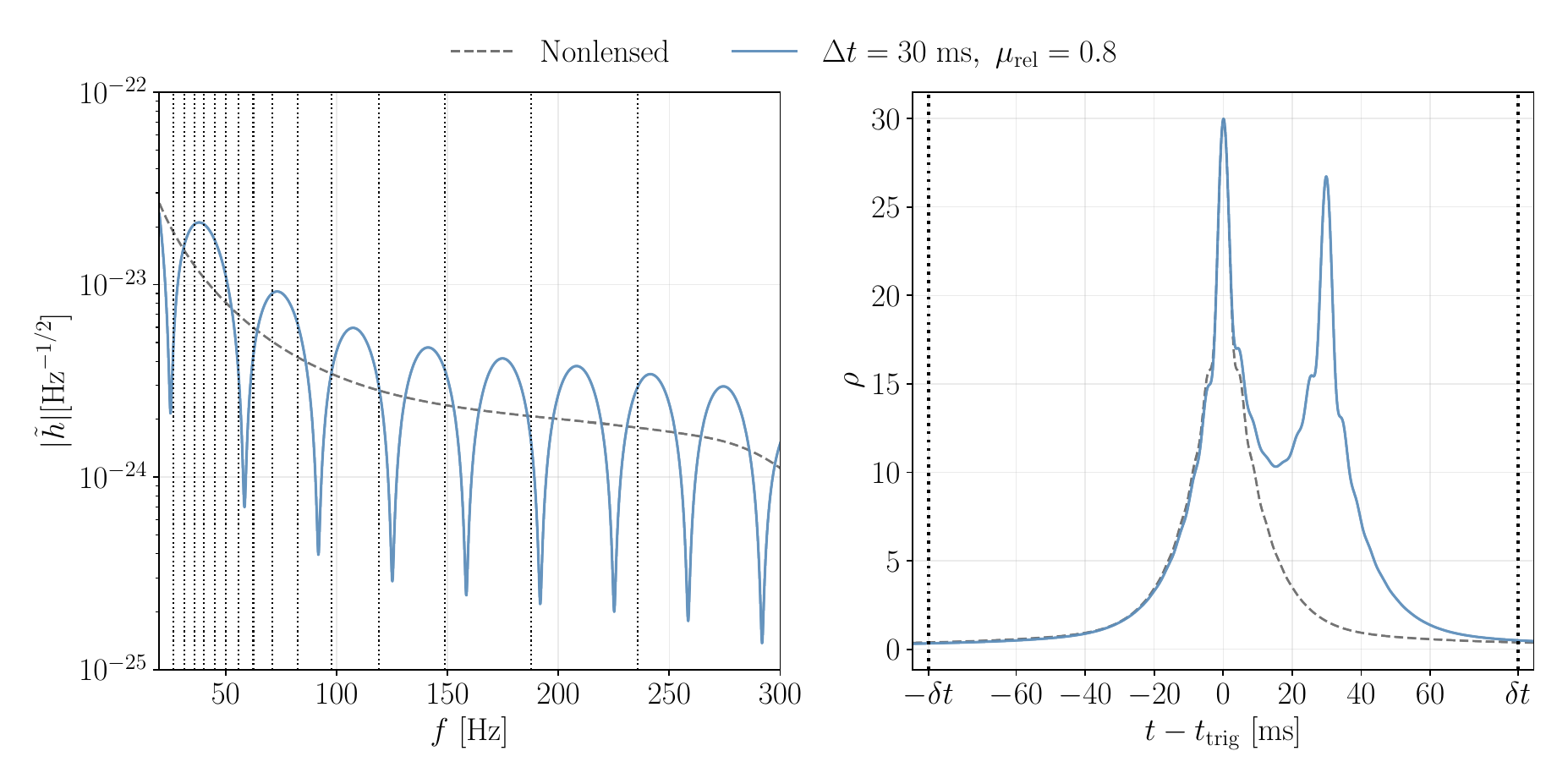}
~    \caption{Schematic impact of lensing on signal-consistency tests for a zero-noise,
two-image $30$--$30\,M_{\odot}$ non-spinning \gls{BBH} signal with $\Delta t=30\,\mathrm{ms}$ and $\mu_{\mathrm{rel}}=0.8$, normalized to a recovered peak \gls{SNR} of $\rho=30$ (solid), compared with a nonlensed reference (dashed).
\textit{Left}: Frequency-domain amplitudes, with vertical lines marking the
$\chi^2$ frequency bins.
\textit{Right}: Matched-filter \gls{SNR} time series as a function of $t-t_{\rm trig}$, with vertical lines marking the signal-consistency window at $\pm85.4\,\mathrm{ms}$.
For the zero-noise example shown here, the first image produces the highest-SNR peak and therefore defines $t_{\rm trig}$.
Lensing-induced interference distorts the expected per-bin \gls{SNR}
contributions and breaks the approximate symmetry of the matched-filter \gls{SNR} time series about $t_{\rm trig}$.}
    \label{fig1}
\end{figure*}

\gls{GW} lensing arises from wave propagation through the curved spacetime generated by an intervening mass distribution.
Under the weak-field and thin-lens approximations, the effect of lensing can be described in the frequency domain through an amplification factor $F(f)$, such that the lensed waveform $\tilde h_{L}(f)$ is related to the nonlensed waveform $\tilde h(f)$ by $\tilde h_{L}(f) = F(f)\tilde h(f)$.
The amplification factor is obtained from the diffraction integral~\citep{Takahashi:2003ix},

\begin{equation} \label{eq:diffraction_integral}
    F(f)=\frac{\tau_D f}{i}\int \mathrm{d}^2\theta\, e^{2\pi i f t_d(\vec\theta,\vec\theta_\mathrm{S})}\,,
\end{equation}

where $t_d(\vec\theta,\vec\theta_\mathrm{S})$ is the lensing time delay for a source at angular position $\vec\theta_\mathrm{S}$ and a position $\vec\theta$ on the lens plane.
Here, $\tau_D=\frac{D_{\rm L} D_{\rm S}}{c D_{\rm LS}}$, where $D_{\rm L}$, $D_{\rm S}$, and $D_{\rm LS}$ are the angular diameter distances from the observer to the lens, from the observer to the source, and between the lens and the source, respectively.
Contributions from different positions on the lens plane acquire different phases through $t_d$, giving rise to diffraction and interference.
The waveform distortions described by $F(f)$ therefore depend on the source--lens--observer configuration and the lens profile.

In the geometric-optics regime, the diffraction integral can be approximated by contributions from discrete stationary points of the time-delay surface.
For two interfering images, after normalizing the amplitude of the first image, the amplification factor can be written as

\begin{equation}\label{eq:two_images_fd}
    F(f)=1+\sqrt{\mu_{\rm rel}}e^{i(2\pi f\Delta t-\pi/2)}
\end{equation}

where $\mu_{\rm rel}$ and $\Delta t$ are the relative magnification and time delay between the two images, respectively, and the $-\pi/2$ term corresponds to their relative Morse phase considered here.
Pairs of dominant images generated by a broad class of lenses, including point-mass lenses and configurations near generic fold caustics~\citep{Blandford:1986zz,Lo:2024wqm}, can therefore be represented by the same interference model parameterized by $\mu_{\rm rel}$, $\Delta t$, and the relative Morse phase.
The coherent superposition produces frequency-domain interference fringes, whose characteristic spacing is $\Delta f=1/\Delta t$, while their modulation strength is controlled by $\mu_{\rm rel}$~\citep{Takahashi:2003ix,Lo:2024wqm,Ezquiaga:2025gkd}, as illustrated in the left panel of Fig.~\ref{fig1}.

The same model has a complementary interpretation in the time domain.
When $\Delta t$ is shorter than the signal duration within the analyzed frequency band, the two images overlap in the strain, but they can remain distinguishable in matched filtering if their separation exceeds the characteristic autocorrelation width of the waveform.
The SNR time series then develops two peaks separated by approximately $\Delta t$, whose relative heights approach $\sqrt{\mu_{\rm rel}}$ for sufficiently resolved images~\citep{Lo:2024wqm,Ezquiaga:2025gkd}.
As shown in the right panel of Fig.~\ref{fig1}, the delayed image also breaks the approximate time symmetry of the nonlensed SNR time series: when the first image defines the trigger, the additional peak and excess SNR appear preferentially on the post-trigger side, while the pre-trigger side remains close to the nonlensed response.

Figure~\ref{fig1} shows the noise-free case.
For $\mu_{\rm rel}\sim1$, however, the two images have comparable SNRs, and detector noise can change which peak is identified as the trigger.
If the delayed image becomes the higher-SNR peak, the additional structure instead lies on the pre-trigger side, reversing the apparent asymmetry about $t_{\rm trig}$.
This ambiguity is discussed further in Sec.~\ref{eta_section}.

\section{The Origin of Lensing Selection Bias}\label{signal_consistency_test}

To examine how lensing affects signal-consistency tests, we consider two representative constructions that probe complementary features of the matched-filter response: the frequency-binned $\chi^2$ test~\citep{Allen:2004gu} and the autocorrelation-based $\xi^2$ test~\citep{Messick:2017a}, which assess consistency across frequency intervals and around the trigger in time, respectively.

Unless otherwise specified, all analyses use the IMRPhenomD waveform model~\citep{Khan:2015jqa}, as implemented in LALSuite~\citep{lalsuite,swiglal}. The analyses are restricted to the frequency range $[20,1024]\,\mathrm{Hz}$, with the detector noise power spectral density corresponding to the third \gls{LVK} observing run~\citep{KAGRA:2021vkt}.

We first consider the $\chi^2$ test introduced
by~\citet{Allen:2004gu}. 
The template is partitioned into $p$ frequency bins, with the boundaries chosen from the noise-weighted template such that each bin contributes equally to the expected matched-filter \gls{SNR}~\citep{Allen:2004gu}.
Let $z_k$ denote the complex matched-filter \gls{SNR} contribution from frequency bin $k$ at the trigger time, and let $z=\sum_k z_k$ denote the total complex \gls{SNR}.
The statistic is
\begin{equation}
    \chi^{2}
    = p\sum_{k=1}^{p}
    \left|z_k-\frac{z}{p}\right|^{2}.
    \label{eq:pycbc_chi2}
\end{equation}

We define the reduced statistic as
$\chi^2_{\rm reduced}=\chi^{2}/(2p-2)$.
We use $p=16$ frequency bins, following the PyCBC configuration in~\citet{Usman:2015kfa}.
The vertical dotted lines in the left panel of Fig.~\ref{fig1} indicate the resulting bin boundaries.
The equal-\gls{SNR} construction yields narrower bins where the nonlensed template has greater noise-weighted power and wider bins where its contribution is smaller.
For a matching signal in Gaussian noise, $\chi^2_{\rm reduced}$ has an expectation value of unity.
A waveform mismatch redistributes the complex matched-filter contribution among the frequency bins, causing the $z_k$ to depart from the expected value $z/p$ and increasing $\chi^2_{\rm reduced}$.
Standard PyCBC ranking statistics therefore begin to down-rank triggers when $\chi^2_{\rm reduced}>1$.
We classify triggers with $\chi^2_{\rm reduced}>10$ as strongly down-ranked, corresponding to $P(\chi^2_{\rm reduced}\geq10)\sim10^{-46}$ for an exactly matching signal in Gaussian noise~\citep{Allen:2004gu}.

The left panel in Fig.~\ref{fig1} illustrates how a lensed \gls{GW} departs from the response expected for a nonlensed template under the $\chi^2$ test.
The lensing delay $\Delta t$ sets the fringe spacing, whereas the frequency-bin boundaries are fixed by the nonlensed template.
The unequal frequency bins therefore generally sample different parts of the interference pattern, causing their complex matched-filter contributions $z_k$ to depart from the expected value $z/p$.
For the illustrated lensed signal, these deviations produce $\chi^2_{\rm reduced}\simeq15$, placing it in the strongly down-ranked regime defined above.
More generally, lensing-induced interference can affect consistency tests that compare matched-filter contributions across frequency regions.
This mechanism may also affect other searches that compare matched-filter contributions across frequency regions, although their constructions can differ.

\begin{figure*}[t]
    \centering
    \includegraphics[width=\textwidth]{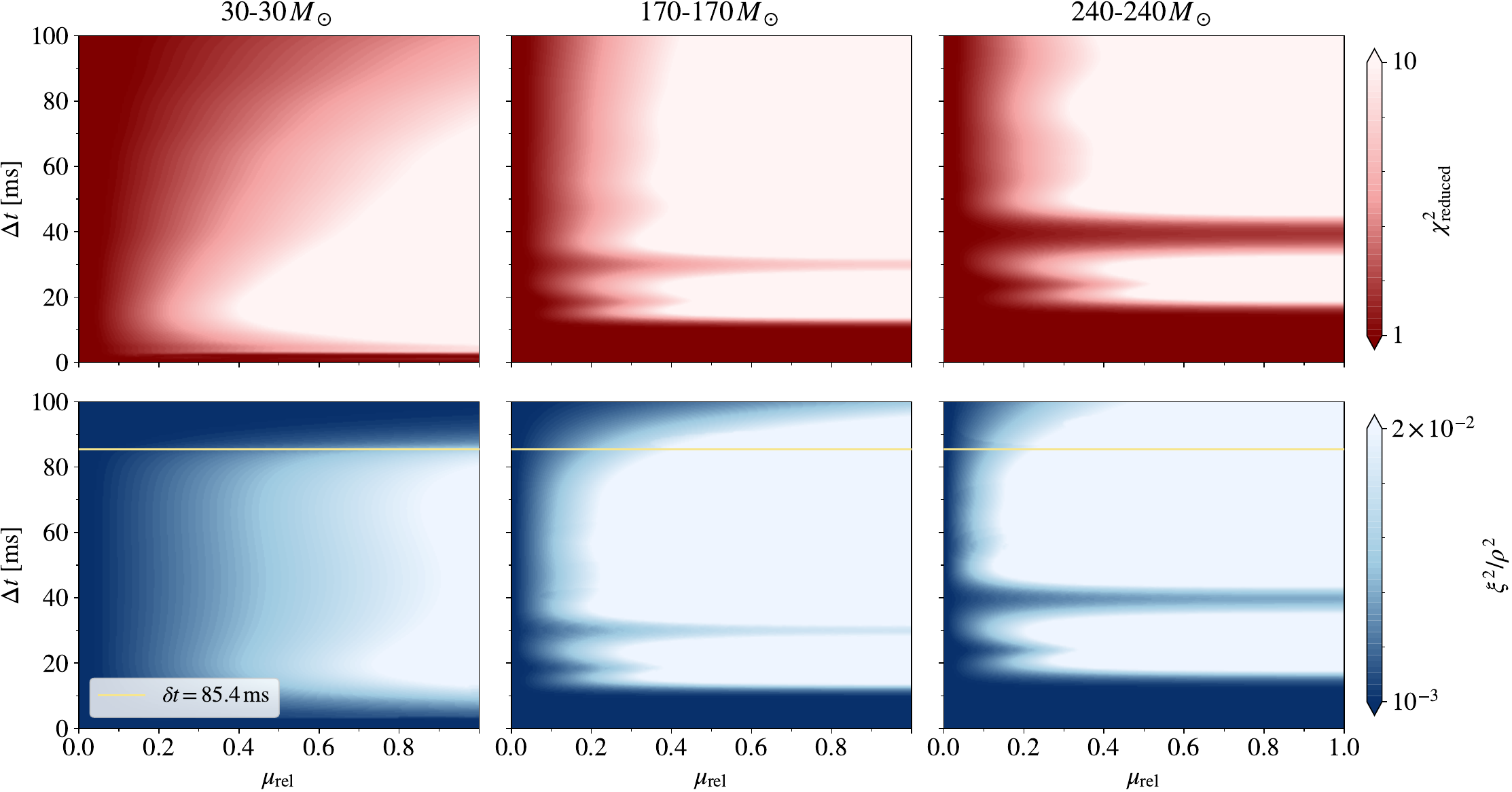}
    \caption{Signal-consistency test values over the two-image lensing parameter ($\mu_{\rm rel}, \Delta t$) space for nonspinning \glspl{BBH} with component masses $30$--$30\,M_\odot$, $170$--$170\,M_\odot$, and $240$--$240\,M_\odot$.
    The signals are normalized to a recovered peak \gls{SNR} of $\rho=30$.
    The top row shows the reduced statistic $\chi^2_{\rm reduced}$, while the bottom row shows the autocorrelation consistency statistic $\xi^2/\rho^2$.
    Regions with $\xi^2/\rho^2 > 2\times10^{-2}$ or $\chi^2_{\rm reduced} > 10$ fail the chosen consistency thresholds. Across all panels, only a limited portion of the lensed parameter space remains consistent under these thresholds.}
    \label{selection-bias}
\end{figure*}

We next consider the $\xi^2$ test, which evaluates the time evolution of the matched-filter response.
For a signal well described by the template, the expected \gls{SNR} time series around the trigger follows the template autocorrelation, with noise producing stochastic residuals and waveform mismatch introducing additional structure.
For a template $h_j$, let $z_j(t)$ denote its complex matched-filter \gls{SNR} time series centered on the trigger, $t_{\rm trig}$, defined by the maximum \gls{SNR}, and let $R_j(t)$ denote the normalized template autocorrelation, with $R_j(0)=1$. Here, $t$ is measured relative to $t_{\rm trig}$.
\begin{equation}
    \xi_j^{2} \propto
    \int_{-\delta t}^{\delta t}
    \left|z_j(t)-z_j(0)R_j(t)\right|^{2}\,\mathrm{d}t ,
    \label{eq:gstlal_xi2}
\end{equation}
where $\delta t\simeq85.4\,\mathrm{ms}$ defines the time window over which signal consistency is evaluated~\citep{Messick:2017a}.
We adopt the normalization of \citet{Messick:2017a}.
For a matching signal in Gaussian noise, the residual remains small, whereas significant waveform distortion within $\delta t$ not captured by the template increases $\xi_j^{2}$.
To compare this inconsistency across triggers with different amplitudes, we consider $\xi^{2}/\rho^{2}$, where $\rho=|z_j(0)|$ is the recovered peak \gls{SNR}.
Triggers with $\xi^{2}/\rho^{2}\gtrsim2\times10^{-2}$ are strongly down-ranked by the search ranking statistic~\citep{Messick:2017a,Magee:2023muf,Chan:2024qmb}.

The right panel of Fig.~\ref{fig1} demonstrates how a delayed image modifies this test.
As discussed in the previous section, when the first image defines \(t_{\rm trig}\), the delayed image produces a localized post-trigger excess while the pre-trigger \gls{SNR} time series remains close to the nonlensed response.
The corresponding contribution to \(\xi^{2}\) is therefore strongly asymmetric about the trigger, while only residuals within the finite window \(|t|\leq\delta t\) contribute to the statistic.
For this example, the distortion yields $\xi^{2}/\rho^{2}\simeq0.5$, well above the $\sim2\times10^{-2}$ scale at which triggers are strongly down-ranked.
This behavior is therefore relevant to other matched-filter searches that compare the \gls{SNR} time series around a trigger with the response expected from the template autocorrelation~\citep{Allene:2025saz,Chu:2020pjv}.
Therefore, while lensing produces broadband structure in the frequency domain, its effect on the $\xi^2$ test is localized around the trigger.

The example in Fig.~\ref{fig1} assumes zero detector noise and a nonlensed template with source parameters identical to those of the lensed signal.
Noise fluctuations and the specific implementation of signal-consistency tests in search pipelines will modify the quantitative values of the statistics, but not the qualitative frequency- and time-domain mechanisms described here.
The degree of down-ranking is therefore inherently pipeline-dependent and must be evaluated within the ranking framework of each search.

\begin{table}[t]
    \centering
    \setlength{\tabcolsep}{4.5pt}
    \renewcommand{\arraystretch}{1.05}
    \begin{tabular}{cc}
        \toprule
        Parameter & Foreground distribution \\
        \midrule
        $\mathcal{M}_c$ [$M_\odot$] & $\mathcal{U}(12,80)$ \\
        $q$ & $\mathcal{U}(0.125,1)$ \\
        $\chi_{1z}$ & $\mathcal{U}(-0.99,0.99)$ \\
        $\chi_{2z}$ & $\mathcal{U}(-0.99,0.99)$ \\
        $\theta_{JN}$ & $\sin(0,\pi)$ \\
        $\phi_{\rm ref}$ & $\mathcal{U}(0,2\pi)$ \\
        $\mathrm{RA}$ & $\mathcal{U}(0,2\pi)$ \\
        $\mathrm{DEC}$ & $\cos(-\pi/2,\,\pi/2)$ \\
        $\psi$ & $\mathcal{U}(0,\pi)$ \\
        $t_c$ [s] & $\mathcal{U}(t_{\min},t_{\max})$ \\
        $d_L$ [Gpc] & $\mathcal{U}(0.1,5)$ \\
        \midrule
        $\mu_{\rm rel}$ & $\mathcal{U}(0,1)$ \\
        $\Delta t$ [s] & $\mathcal{U}(0,0.1)$ \\
        \bottomrule
    \end{tabular}
    \caption{Source and lensing parameter proposal distributions used
    for the foreground analysis.
    Here, $\mathcal{M}_c$ is the detector-frame chirp mass,
    $q=m_2/m_1\leq1$ is the mass ratio, and $\chi_{iz}$ are the
    dimensionless spin components along the orbital angular momentum.
    The viewing angle $\theta_{JN}$ is measured between the total
    angular momentum and the line of sight; $\phi_{\rm ref}$ is the
    orbital phase at the reference frequency.
    RA and DEC specify the sky position, $\psi$ is the polarization
    angle, $t_c$ is the geocentric coalescence GPS time, and $d_L$ is the
    luminosity distance. All angles are in radians.
    $\mathcal{U}$ denotes a uniform distribution.
    The sine and cosine entries denote densities proportional to
    $\sin\theta_{JN}$ and $\cos(\mathrm{DEC})$ on the stated intervals.
    The signed spin components are drawn independently;
    transverse components vanish.
    The angular distributions have support
    $\theta_{JN}\in[0,\pi]$ and
    $\mathrm{DEC}\in[-\pi/2,\pi/2]$.
    The coalescence GPS time is drawn between
    $t_{\min}=1238173083$ and $t_{\max}=1238785101$.}
    \label{tab:waveform_settings}
\end{table}
\section{Parameter Dependence and Local Minima}\label{sec:selection-bias}

We next characterize the lensing-induced changes in these representative consistency tests beyond the single configuration.
Figure~\ref{selection-bias} maps the consistency between lensed signals and nonlensed templates over the $(\mu_{\rm rel},\Delta t)$ parameter space for three representative equal-mass systems.
The dependence on $\mu_{\rm rel}$ is similar for both statistics, whereas their behavior with $\Delta t$ differs.
The inconsistency generally increases with $\mu_{\rm rel}$, as stronger two-image interference produces larger deviations from the nonlensed template.
For $\chi^2_{\rm reduced}$, the inconsistency decreases once the interference fringes become sufficiently fine to average out within the frequency bins.
In contrast, $\xi^2/\rho^2$ exhibits a more localized drop at finite $\Delta t$, set by the autocorrelation structure of the signal.
While an analogous sensitivity study is not yet available to validate the $\chi^2_{\rm reduced}$ behavior in an end-to-end search, the $\xi^2/\rho^2$ trend is qualitatively consistent with a recent point-mass lensing sensitivity study, which found this statistic to dominate the loss in search sensitivity~\citep{Chan:2024qmb}.
This agreement further supports the relevance of the behavior identified here to realistic search selection effects, as illustrated in Appendix~\ref{apx:pml}.

The inconsistent region also expands with increasing total mass, although for different reasons in the two tests.
For $\chi^2_{\rm reduced}$, the narrower bandwidth of high-mass signals requires a smaller fringe spacing for the lensing-induced SNR residual to average out within the frequency bins.
For $\xi^2/\rho^2$, the narrower bandwidth produces a broader autocorrelation function, allowing the lensing-induced SNR residual to persist over a wider range of image delays.
Both statistics develop a local minimum at finite $\Delta t$ for the higher-mass systems, restricting the distorted-image phenomenology that can remain sufficiently consistent with nonlensed templates to a narrow region of parameter space.
This behavior is relevant to GW231123, a high-mass, highly
spinning \gls{BBH} event~\citep{LVK:2025b} for which several
analyses report a preference for wave-optics lensing
models~\citep{LVKlensing:2025cwb,Chan:2025kyu,
Goyal:2025eqo,Cheung:2026pky,Wang:2026yjk}.
Its lensing interpretation remains unconfirmed, with
waveform systematics and astrophysical prior assumptions
affecting the assessment~\citep{LVKlensing:2025cwb,
Chan:2025kyu,Cheung:2026pky}.
We examine its signal-consistency behavior in
Appendix~\ref{apx:GW231123}.

Interference occurs across a broad class of lens models, making these effects relevant beyond the specific configurations considered here.
For the systems and thresholds examined here, broad regions of the two-image parameter space produce strong signal inconsistency.
Such consistency penalties can exclude otherwise significant signals from standard matched-filter detections~\citep{Chan:2024qmb}.
All distorted-image analyses based on these detections therefore inherit lensing-dependent selection effects, which can substantially suppress the representation of strongly distorted signals in the detected events.
An approach that searches for distorted images only after standard matched-filter detection, while neglecting these selection effects, is therefore insufficient for robust lensing-rate estimates and lens-population constraints.
These results establish the need to account for lensing in search sensitivity and motivate using its structured residuals to improve the recovery of distorted signals.

\section{Turning Lensing Residuals into a New Search Statistic}\label{eta_section}

\begin{figure}[t]
    \includegraphics[width=\columnwidth]{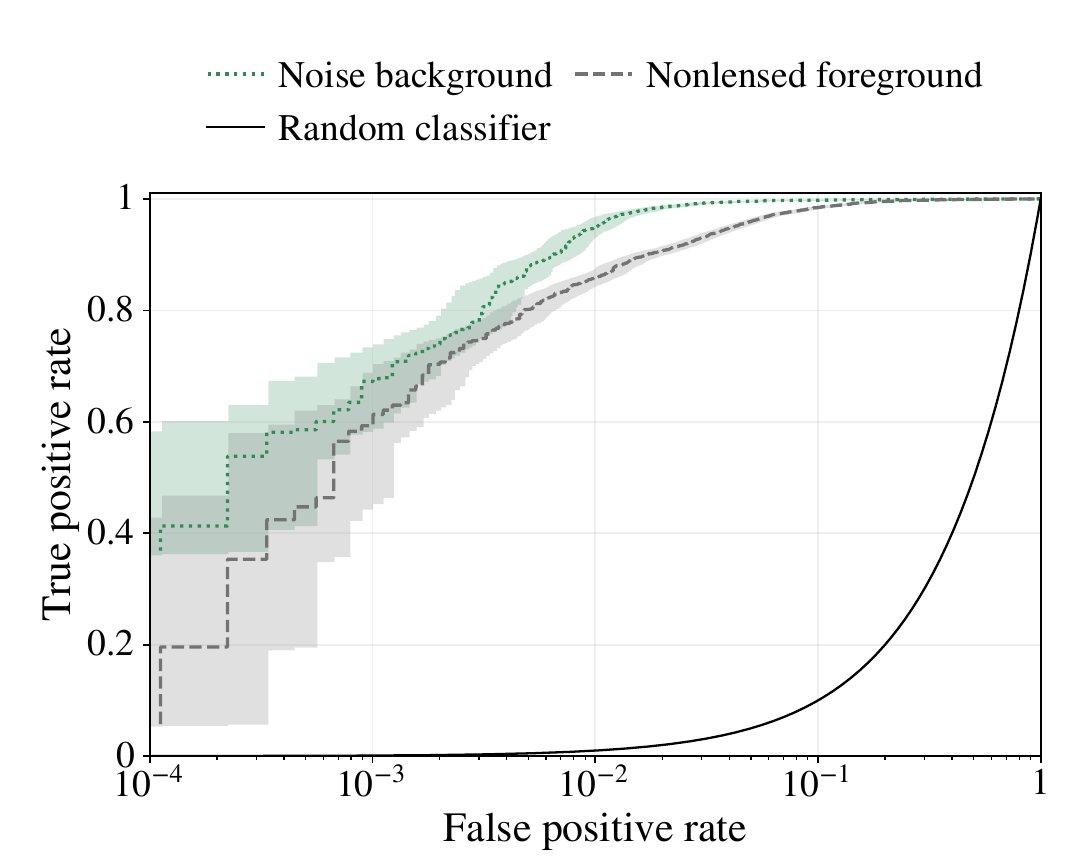}
    \caption{  \gls{ROC} curves for $\eta$, using $\sim9000$ selected lensed foreground injections as the positive population and classifying events as
  positive when $\eta\geq\eta_{\rm th}$.
  The comparison populations contain the same number of noise-background triggers (green dotted; \gls{AUC} $=0.9972^{+0.0006}_{-0.0006}$) and
  paired nonlensed foreground injections (gray dashed; \gls{AUC} $=0.9917^{+0.0009}_{-0.0010}$).
  Shaded bands show $95\%$ confidence intervals.
  The black solid curve denotes random classification, for which the true-positive and false-positive rates are equal.}
    \label{eta_plot}
\end{figure}

To mitigate the selection effects that follow-up analyses inherit from detection searches, previously suggested strategies include lensing follow-up of candidates identified by coherent WaveBurst and extending matched-filter template banks to include lensing~\citep{Chan:2024qmb}.
For searches using the $\xi^2$ test considered here, we further identify $0\leq\Delta t\leq\delta t$ and $0\leq\mu_{\rm rel}\leq1$ as a target region for a two-image template bank applicable to a broad class of lens models.
However, extending the template bank to include lensing may be costly, since computational cost can grow exponentially with search dimensionality.
We therefore exploit the asymmetric \gls{SNR} residuals to construct an additional search statistic aimed at recovering severely down-ranked lensed \glspl{GW} without expanding the template bank.

As discussed in Sec.~\ref{signal_consistency_test}, noise fluctuations can change which of two comparable \gls{SNR} peaks defines the trigger.
We therefore define an asymmetry statistic invariant under exchange of the contributions before and after the trigger,

\begin{equation}
\eta =
\frac{\max(\xi^2_{\rm l},\xi^2_{\rm r})}
{\min(\xi^2_{\rm l},\xi^2_{\rm r})},
\label{eta_def}
\end{equation}

where $\xi^2_{\rm l}$ and $\xi^2_{\rm r}$ are the contributions to $\xi^2$ evaluated over the intervals $[-\delta t,0]$ and $[0,\delta t]$, respectively.
By construction, $\eta\geq1$, with $\eta=1$ indicating equal residual power before and after the trigger.
A large $\eta$ identifies residual power concentrated on one side of the trigger, independent of whether the excess occurs before or after the trigger.

To evaluate the performance of $\eta$, we construct paired injections consisting of a nonlensed signal and its lensed counterpart.
The source and lensing parameters are sampled from the distributions summarized in Table~\ref{tab:waveform_settings}.
The injections are added to real detector data.
Nonlensed injections satisfy $7<\rho<30$ and $\xi^2/\rho^2<0.01$, while their paired lensed counterparts satisfy $\rho>7$ and $\xi^2/\rho^2>0.02$.
Only pairs satisfying both selections are retained.
This selection focuses on signals that retain $\rho>7$ but can be strongly down-ranked because of lensing-induced inconsistency.
We compare their $\eta$ values with those of the nonlensed counterparts and background triggers obtained from the nonlensed template-bank injection campaign described in Appendix~\ref{apx:pml}.

We quantify the performance using the \gls{ROC} curves in Fig.~\ref{eta_plot}, evaluating the fraction of selected lensed injections retained against the fraction retained from each comparison population as the threshold on $\eta$ is varied.
The \gls{AUC} values are $0.9972$ against noise-background triggers and $0.9917$ against nonlensed injections, indicating strong separation from both populations within the selected sample.
Although further studies are needed to establish whether incorporating $\eta$ improves search sensitivity at fixed false-alarm rate, these results identify $\eta$ as a promising candidate for improving the recovery of down-ranked lensed signals without expanding the template bank.

\section{Conclusion}\label{Section: Conclusion}

We identify a shared mechanism for lensing-induced selection effects across lens models and matched-filtering pipelines.
The distorted signal leaves \gls{SNR} residuals relative to the response expected from a nonlensed template in both signal-consistency tests and can therefore be substantially down-ranked in addition to the reduction in matched-filter \gls{SNR}.
In the time domain, the delayed image can produce a localized and asymmetric residual around the trigger time, whereas in the frequency domain the interference fringes alter matched-filter contributions across multiple frequency bins.
For higher-mass templates, both statistics develop local minima at finite $\Delta t$, identifying regions where lensing-induced consistency penalties are reduced.

Lensing-induced consistency penalties become more pronounced at higher \gls{SNR}, so even loud lensed signals can be strongly down-ranked~\citep{Chan:2024qmb}.
Distorted-image analyses based on matched-filter detections can therefore inherit biases in the source and lensing parameter space, affecting lens-population constraints and lensing-rate estimates.
These results highlight the need to account for lensing-induced inconsistency in search strategies rather than relying on the recovered \gls{SNR} alone.

We turn the residuals responsible for lensing-induced down-ranking into an opportunity to construct a new search statistic, $\eta$, that exploits their asymmetric time-domain morphology.
Within the selected injection sample, $\eta$ achieves \gls{AUC} values of $0.9972$ against noise-background triggers and $0.9917$ against paired nonlensed injections, demonstrating strong separation from both populations.
Whether $\eta$ improves search sensitivity at remains to be established, but these results highlight its potential as a search statistic.

A natural next step is to establish the potential search sensitivity gain against the noise background with mathched-filtering search pipeline.
Once validated, the modified search could be applied to data from previous observing runs through re-filtering, providing an opportunity to recover lensed signals previously down-ranked by conventional signal-consistency tests.

Although our analysis focuses on lensing-induced selection effects, the same approach could be applied to investigate signal-consistency penalties caused by overlapping signals in next-generation ground-based detectors~\citep{Pizzati:2021apa,Alvey:2023naa,Baka:2025yqx}.

\clearpage
\appendix

\section{Consistency with Matched-filtering Search Results}\label{apx:pml}

We reexamine a point-mass-lens matched-filtering injection campaign from \citet{Chan:2024qmb} to compare the time-domain selection effect identified in Sec.~\ref{sec:selection-bias} with the behavior of an end-to-end search.
The campaign consists of nonspinning $30$--$30\,M_\odot$ binary black holes at a fixed luminosity distance of $125\,\mathrm{Mpc}$ analyzed with the \gstlal{} search pipeline.
We refer to \citet{Chan:2024qmb} for the injection setup and the construction of the ranking statistic and false-alarm rate.

To compare these injections directly with the phenomenological lens parameters used in Sec.~\ref{sec:selection-bias}, we transform the point-mass parameters $(M_{\rm Lz},y)$, where $M_{\rm Lz}$ is the redshifted lens mass and $y$ is the source position in units of the Einstein angle, to the relative time delay $\Delta t$ and relative magnification $\mu_{\rm rel}$~\citep{Takahashi:2003ix},

\begin{align}
    \Delta t &=
    \frac{4GM_{\mathrm{Lz}}}{c^3}\Delta T(y),
    \label{eqn:pml_dt}\\
    \mu_{\rm rel} &=
    \left|
    \frac{
    y\sqrt{y^2+4}-(y^2+2)
    }{
    y\sqrt{y^2+4}+(y^2+2)
    }
    \right|,
    \label{eqn:pml_murel}
\end{align}

where

\begin{equation}
    \Delta T(y)
    =
    \frac{1}{2}y\sqrt{y^2+4}
    + \ln\left|
    \frac{\sqrt{y^2+4}+y}
         {\sqrt{y^2+4}-y}
    \right|.
    \label{eqn:pml_DeltaT}
\end{equation}

Here, $\Delta t$ scales linearly with $M_{\rm Lz}$, while
$\Delta T$ and $\mu_{\rm rel}$ depend only on $y$.

\begin{figure}[t]
    \includegraphics[width=0.9\columnwidth]
    {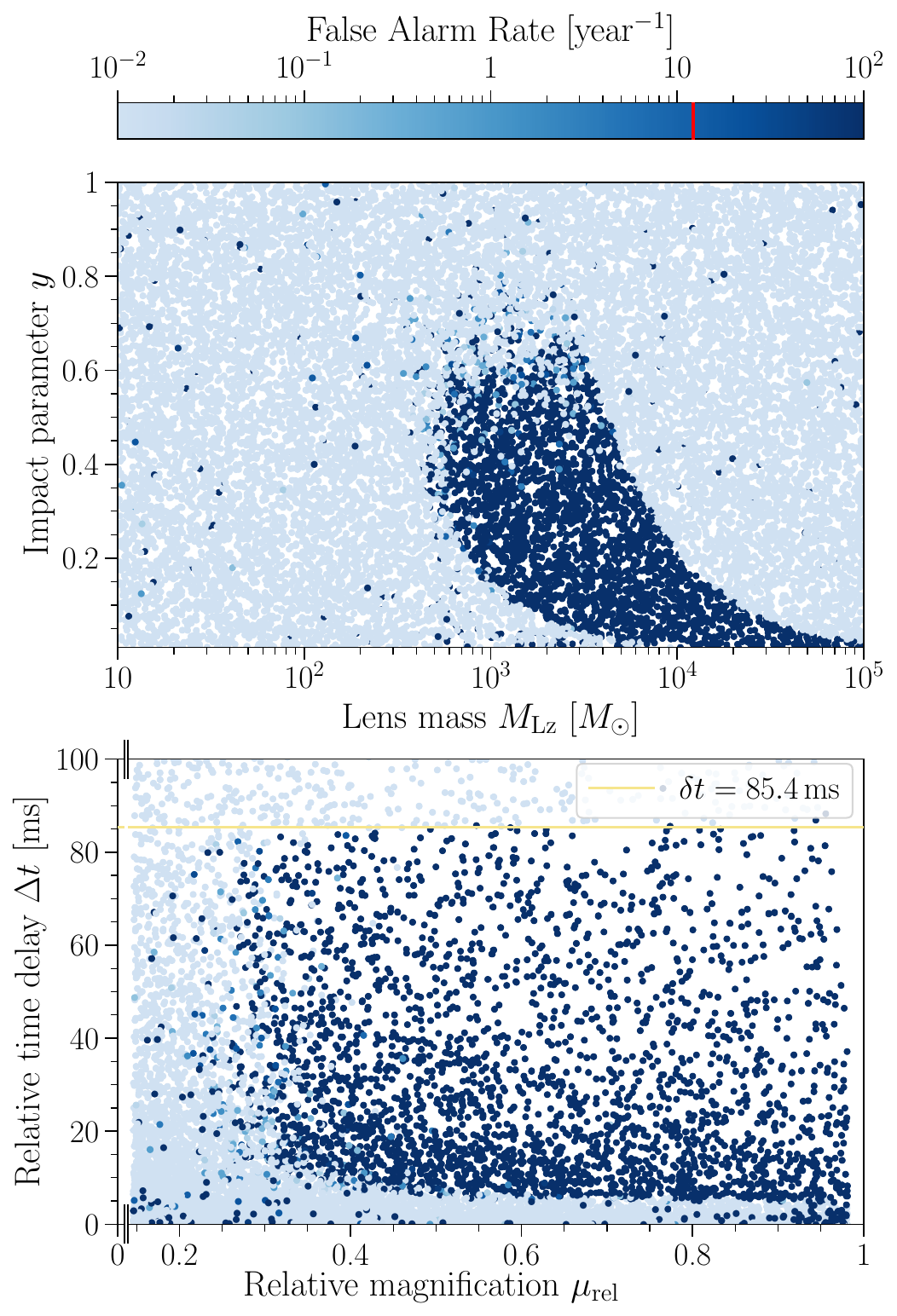}
    \caption{
    Upper: Point-mass-lens injections campaign result in $(M_{\rm Lz},y)$ color-coded
    by false-alarm rate from~\cite{Chan:2024qmb}. Lower: The same injections reparameterized in
    $(\mu_{\rm rel},\Delta t)$ using
    Eqs.~\eqref{eqn:pml_dt}--\eqref{eqn:pml_DeltaT}.
    The red marker indicate the false-alarm rate threshold of detection in 
    The yellow line marks the $\delta t=85.4\,\mathrm{ms}$ window used
    by $\xi^2$ test.
    The lost injections are concentrated within the $\delta t$ window, where the second image temporally overlaps with the first and affects the $\xi^2$ test.
    }
    \label{gstlal_search_result}
\end{figure}

Figure~\ref{gstlal_search_result} shows the injections in
$(M_{\rm Lz},y)$ and reparameterized in $(\mu_{\rm rel},\Delta t)$,
with points color-coded by FAR. A broad region of lensing parameter
space is down-ranked by the search, with the effect concentrated
predominantly at $\Delta t<85.4\,\mathrm{ms}$, the time window used by
the \gstlal{} autocorrelation-based $\xi^2$ test. This is consistent
with the localized dependence on $\Delta t$ found in
Sec.~\ref{sec:selection-bias}, where overlapping lensed images perturb
the expected matched-filter autocorrelation within the same window.

This effect arises from temporally overlapping images rather than the specific point-mass parameterization, suggesting qualitatively similar behavior for other lens models that produce comparable image delays and magnifications. 
Including lensed templates covering this region of parameter space should therefore reduce the associated selection bias by reducing the mismatch that drives the signal-consistency penalty. 
Comparable end-to-end injection campaigns are not available here for frequency-binned $\chi^2$ tests used by other matched-filtering searches, so this comparison supports specifically the localized time-domain effect associated with the \gstlal{} $\xi^2$ test.

\section{Signal Consistency under the Lensing Hypothesis for GW231123}\label{apx:GW231123}

\begin{table}[t]
    \centering
    \setlength{\tabcolsep}{6pt}
    \renewcommand{\arraystretch}{1.05}
    \begin{tabular}{cc}
        \toprule
        Parameter & Value \\
        \midrule
        $\mathcal{M}_c$ [$M_\odot$] & 107.209 \\
        $q$ & 0.321674 \\
        $m_1$ [$M_\odot$] & 223.875 \\
        $m_2$ [$M_\odot$] & 72.0149 \\
        $a_1$ & 0.734566 \\
        $a_2$ & 0.736988 \\
        $\phi_{12}$ & 2.301013 \\
        $\theta_1$ & 0.623115 \\
        $\theta_2$ & 1.336449 \\
        $\phi_{JL}$ & 1.076094 \\
        $\theta_{JN}$ & 2.800685 \\
        $\phi_{\rm ref}$ & 2.116213 \\
        $\mathrm{RA}$ & 2.864087 \\
        $\mathrm{DEC}$ & -0.121224 \\
        $\psi$ & 1.562047 \\
        $t_c$ [s] & 1384782888.617595 \\
        $d_L$ [Gpc] & 3.722869 \\
        \midrule
        $\mu_{\rm rel}$ & $[0,1]$ \\
        $\Delta t$ [s] & $[0,0.1]$ \\
        \bottomrule
    \end{tabular}
    \caption{
    Source parameters of the maximum-likelihood lensed sample for
    GW231123 used to generate the injected signal in the
    signal-consistency calculation. The lensing rows give the
    scanned ranges, rather than maximum-likelihood values.
    Shared parameter definitions and angular units follow
    Table~\ref{tab:waveform_settings}.
    The component masses $m_i$ are in the detector frame, $a_i$ are
    the dimensionless spin magnitudes, and $\theta_i$ are the spin
    tilt angles relative to the orbital angular momentum.
    The angle $\phi_{12}$ is the difference between the spin
    azimuths, and $\phi_{JL}$ is the azimuth of the orbital angular
    momentum about the total angular momentum.
    }
    \label{tab:GW231123_injection}
\end{table}

\begin{figure}[t]
    \includegraphics[width=\columnwidth]{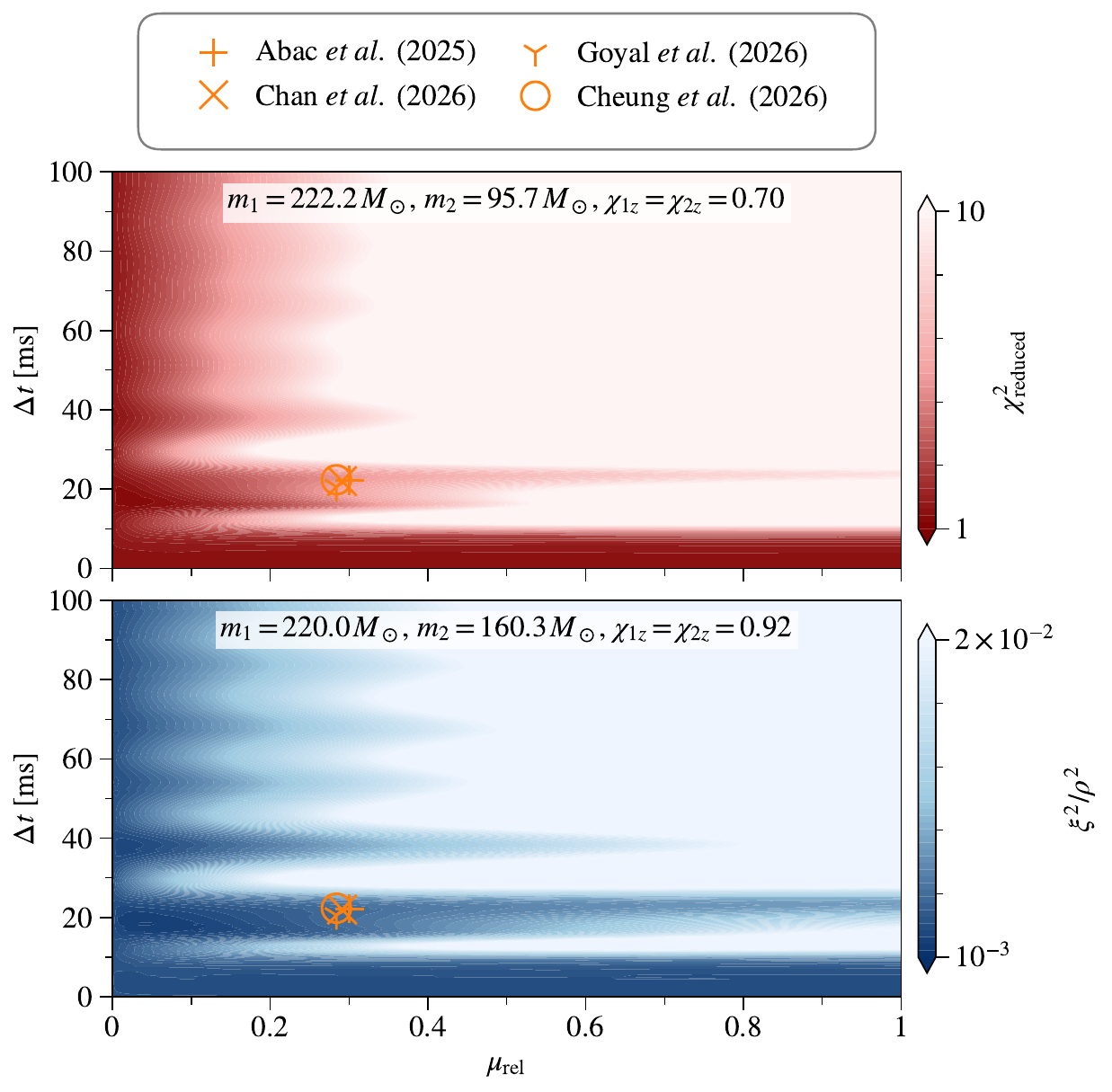}
    \caption{
    Signal-consistency statistics evaluated using the intrinsic
    parameters of the templates recovered for GW231123 by PyCBC
    (upper) and \gstlal{} (lower), as functions of
    $(\mu_{\rm rel},\Delta t)$. Orange markers show recent lensing
    parameter estimates for GW231123; point-mass-lens results are
    transformed using Eqs.~\eqref{eqn:pml_dt}--%
    \eqref{eqn:pml_DeltaT}. Both statistics exhibit a local minimum
    near $\Delta t\sim20\,\mathrm{ms}$, close to the inferred lensing
    parameters.
    }
    \label{GW231123}
\end{figure}

High-mass \glspl{BBH} are generally subject to stronger
lensing-induced selection effects, but the local minima in the
$\xi^2$ and $\chi^2$ statistics identified in
Sec.~\ref{sec:selection-bias} provide restricted regions where the
signal-consistency penalty is reduced.
GW231123 provides a useful case study of this behavior
because it is a high-mass event recovered by both
\gstlal{} and PyCBC~\citep{LVK:2025b}.

The template recovered by a practical matched-filtering search need not coincide with the optimal nonlensed approximation adopted in the controlled calculations of Sec.~\ref{sec:selection-bias}.
We therefore repeat the analysis using the publicly available template parameters recovered for GW231123 by each search pipeline~\citep{LVK:2025a}.
For the injected lensed signal, we adopt the maximum-likelihood sample from the GW231123 lensing analysis of \citet{Chan:2025kyu}, with the source parameters summarized in Table~\ref{tab:GW231123_injection}.
For each pipeline, we construct the corresponding nonlensed recovered template and evaluate its signal-consistency statistic against the lensed NRSur7dq4 signal~\citep{Varma:2019csw} over $(\mu_{\rm rel},\Delta t)$, using the publicly released H1 noise power spectral density~\citep{LVK:2025c}.
This calculation isolates the response of the recovered search template to lensing-induced waveform distortions rather than reproducing the full search ranking.

Figure~\ref{GW231123} shows that both statistics retain the qualitative structure found in Sec.~\ref{sec:selection-bias}.
In particular, the high-mass templates recovered for GW231123 exhibit a local minimum near $\Delta t\sim20\,\mathrm{ms}$ in both $\xi^2/\rho^2$ and
$\chi^2_{\rm reduced}$.
We also show the lensing parameters inferred for GW231123 in recent analyses~\citep{LVKlensing:2025cwb,Chan:2025kyu,
Goyal:2025eqo,Cheung:2026pky}.
Results expressed in the point-mass lens parameters $(M_{\rm Lz},y)$ are converted to
$(\mu_{\rm rel},\Delta t)$ using Eqs.~\eqref{eqn:pml_dt}--\eqref{eqn:pml_DeltaT}.
The inferred parameters cluster near the reduced-penalty region around $\Delta t\sim20\,\mathrm{ms}$.

The persistence of this minimum for the templates actually recovered by both searches demonstrates that it is not specific to the idealized template configurations considered in Sec.~\ref{sec:selection-bias}.
More generally, wave-optics distortions of high-mass templates can lead to larger signal-consistency penalties across much of the lensing parameter space, while localized minima identify restricted regions in which the same signals are less strongly down-ranked.
GW231123 lies within such a region and is therefore compatible with the presence of a severe lensing selection effect.

\begin{acknowledgements}
We are grateful to Otto Hannuksela and Alvin K.Y. Li for useful comments and discussions along this project.
We also thank Alvin K.Y. Li for the internal LVK review.
The authors acknowledge the use of OpenAI's ChatGPT to assist with proofreading, language editing, and improving the clarity and structure of the manuscript. 
All AI-assisted content was reviewed and verified by the authors, who take full responsibility for the accuracy and integrity of the final manuscript.
E.K.Y.L. acknowledges support from the Research Grants Council of Hong Kong (Project Nos.\ CUHK 14304622, 14307923, and 14307724), as well as from the Direct Grant for Research awarded by the Research Committee of The Chinese University of Hong Kong.
This research has made use of data or software obtained from the Gravitational Wave Open Science Center (gwosc.org), a service of the LIGO Scientific Collaboration, the Virgo Collaboration, and KAGRA. This material is based upon work supported by NSF's LIGO Laboratory which is a major facility fully funded by the National Science Foundation, as well as the Science and Technology Facilities Council (STFC) of the United Kingdom, the Max-Planck-Society (MPS), and the State of Niedersachsen/Germany for support of the construction of Advanced LIGO and construction and operation of the GEO600 detector. Additional support for Advanced LIGO was provided by the Australian Research Council. Virgo is funded, through the European Gravitational Observatory (EGO), by the French Centre National de Recherche Scientifique (CNRS), the Italian Istituto Nazionale di Fisica Nucleare (INFN) and the Dutch Nikhef, with contributions by institutions from Belgium, Germany, Greece, Hungary, Ireland, Japan, Monaco, Poland, Portugal, Spain. KAGRA is supported by Ministry of Education, Culture, Sports, Science and Technology (MEXT), Japan Society for the Promotion of Science (JSPS) in Japan; National Research Foundation (NRF) and Ministry of Science and ICT (MSIT) in Korea; Academia Sinica (AS) and National Science and Technology Council (NSTC) in Taiwan.
\end{acknowledgements}

\bibliographystyle{aasjournalv7.1}
\bibliography{citations}

\begin{thebibliography}{}
\expandafter\ifx\csname natexlab\endcsname\relax\def\natexlab#1{#1}\fi
\providecommand{\url}[1]{\href{#1}{#1}}
\providecommand{\dodoi}[1]{doi:~\href{http://doi.org/#1}{\nolinkurl{#1}}}
\providecommand{\doeprint}[1]{\href{http://ascl.net/#1}{\nolinkurl{http://ascl.net/#1}}}
\providecommand{\doarXiv}[1]{\href{https://arxiv.org/abs/#1}{\nolinkurl{https://arxiv.org/abs/#1}}}

\bibitem[{A.~G. Abac {et~al.}(2025{\natexlab{a}})Abac {et~al.}}]{LVKlensing:2025cwb}
Abac, A.~G., {et~al.} 2025{\natexlab{a}}, \bibinfo{title}{{GWTC-4.0: Searches for Gravitational-Wave Lensing Signatures},} \doarXiv{2512.16347}

\bibitem[{A.~G. Abac {et~al.}(2025{\natexlab{b}})Abac {et~al.}}]{LVK:2025b}
Abac, A.~G., {et~al.} 2025{\natexlab{b}}, \bibinfo{title}{{GW231123: a Binary Black Hole Merger with Total Mass 190-265 $M_{\odot}$},} \doarXiv{2507.08219}

\bibitem[{R. Abbott {et~al.}(2023)Abbott {et~al.}}]{KAGRA:2021vkt}
Abbott, R., {et~al.} 2023, \bibinfo{title}{{GWTC-3: Compact Binary Coalescences Observed by LIGO and Virgo during the Second Part of the Third Observing Run},} Phys. Rev. X, 13, 041039, \dodoi{10.1103/PhysRevX.13.041039}

\bibitem[{R. Abbott {et~al.}(2024)Abbott {et~al.}}]{LVKlensing:2023bwz}
Abbott, R., {et~al.} 2024, \bibinfo{title}{{Search for Gravitational-lensing Signatures in the Full Third Observing Run of the LIGO{\textendash}Virgo Network},} Astrophys. J., 970, 191, \dodoi{10.3847/1538-4357/ad3e83}

\bibitem[{B. Allen(2005)Allen}]{Allen:2004gu}
Allen, B. 2005, \bibinfo{title}{{${\chi}^{2}$ time-frequency discriminator for gravitational wave detection},} Phys. Rev. D, 71, 062001, \dodoi{10.1103/PhysRevD.71.062001}

\bibitem[{B. Allen {et~al.}(2012)Allen, Anderson, Brady, Brown, \& Creighton}]{Allen:2005fk}
Allen, B., Anderson, W.~G., Brady, P.~R., Brown, D.~A., \& Creighton, J. D.~E. 2012, \bibinfo{title}{{FINDCHIRP: An Algorithm for detection of gravitational waves from inspiraling compact binaries},} Phys. Rev. D, 85, 122006, \dodoi{10.1103/PhysRevD.85.122006}

\bibitem[{C. All{\'e}n{\'e} {et~al.}(2025)All{\'e}n{\'e} {et~al.}}]{Allene:2025saz}
All{\'e}n{\'e}, C., {et~al.} 2025, \bibinfo{title}{{The MBTA pipeline for detecting compact binary coalescences in the fourth LIGO-Virgo-KAGRA observing run},} Class. Quant. Grav., 42, 105009, \dodoi{10.1088/1361-6382/add234}

\bibitem[{J. Alvey {et~al.}(2023)Alvey, Bhardwaj, Nissanke, \& Weniger}]{Alvey:2023naa}
Alvey, J., Bhardwaj, U., Nissanke, S., \& Weniger, C. 2023, \bibinfo{title}{{What to do when things get crowded? Scalable joint analysis of overlapping gravitational wave signals},} \doarXiv{2308.06318}

\bibitem[{F. Aubin {et~al.}(2021)Aubin {et~al.}}]{Aubin:2020goo}
Aubin, F., {et~al.} 2021, \bibinfo{title}{{The MBTA pipeline for detecting compact binary coalescences in the third LIGO{\textendash}Virgo observing run},} Class. Quant. Grav., 38, 095004, \dodoi{10.1088/1361-6382/abe913}

\bibitem[{T. Baka {et~al.}(2025)Baka, Narola, Janquart, Samajdar, Dietrich, \& Van Den~Broeck}]{Baka:2025yqx}
Baka, T., Narola, H., Janquart, J., {et~al.} 2025, \bibinfo{title}{{Overlapping signals in next-generation gravitational wave observatories: A recipe for selecting the best parameter estimation technique},} Phys. Rev. D, 112, 082001, \dodoi{10.1103/8cwp-mxcd}

\bibitem[{A. Barsode(2026)Barsode}]{Barsode:2026bcs}
Barsode, A. 2026, \bibinfo{title}{{Tool for accurate model selection using Bayes factors},} Phys. Rev. D, 114, 064045, \dodoi{10.1103/wv11-h53b}

\bibitem[{R. Blandford {\&} R. Narayan(1986)Blandford \& Narayan}]{Blandford:1986zz}
Blandford, R., \& Narayan, R. 1986, \bibinfo{title}{{Fermat's principle, caustics, and the classification of gravitational lens images},} Astrophys. J., 310, 568, \dodoi{10.1086/164709}

\bibitem[{K. {Cannon} {et~al.}(2021){Cannon}, {Caudill}, {Chan}, {Cousins}, {Creighton}, {Ewing}, {Fong}, {Godwin}, {Hanna}, {Hooper}, {Huxford}, {Magee}, {Meacher}, {Messick}, {Morisaki}, {Mukherjee}, {Ohta}, {Pace}, {Privitera}, {de Ruiter}, {Sachdev}, {Singer}, {Singh}, {Tapia}, {Tsukada}, {Tsuna}, {Tsutsui}, {Ueno}, {Viets}, {Wade}, \& {Wade}}]{Cannon:2021a}
{Cannon}, K., {Caudill}, S., {Chan}, C., {et~al.} 2021, \bibinfo{title}{{GstLAL: A software framework for gravitational wave discovery},} SoftwareX, 14, 100680, \dodoi{10.1016/j.softx.2021.100680}

\bibitem[{A. Chakraborty {\&} S. Mukherjee(2024)Chakraborty \& Mukherjee}]{Chakraborty:2024mbr}
Chakraborty, A., \& Mukherjee, S. 2024, \bibinfo{title}{{$\mu$-GLANCE: A Novel Technique to Detect Chromatically and Achromatically Lensed Gravitational Wave Signals},} \doarXiv{2410.06995}

\bibitem[{A. Chakraborty {\&} S. Mukherjee(2025)Chakraborty \& Mukherjee}]{Chakraborty:2025maj}
Chakraborty, A., \& Mukherjee, S. 2025, \bibinfo{title}{{The First Model-independent Chromatic Microlensing Search: No Evidence in the Gravitational Wave Catalog of LIGO{\textendash}Virgo{\textendash}KAGRA},} Astrophys. J., 990, 68, \dodoi{10.3847/1538-4357/adf330}

\bibitem[{A. Chakraborty {\&} S. Mukherjee(2026)Chakraborty \& Mukherjee}]{Chakraborty:2025pxt}
Chakraborty, A., \& Mukherjee, S. 2026, \bibinfo{title}{{The First Model-independent Upper Bound on Microlensing Signature of the Highest-mass Binary Black Hole Event GW231123},} Astrophys. J., 1003, 20, \dodoi{10.3847/1538-4357/ae61a7}

\bibitem[{J.~C.~L. Chan {et~al.}(2026)Chan, Ezquiaga, Lo, Bowman, Maga{\~n}a~Zertuche, \& Vujeva}]{Chan:2025kyu}
Chan, J. C.~L., Ezquiaga, J.~M., Lo, R. K.~L., {et~al.} 2026, \bibinfo{title}{{Discovering gravitational waveform distortions from lensing: A deep dive into GW231123},} Phys. Rev. D, 114, L021502, \dodoi{10.1103/5292-948q}

\bibitem[{J.~C.~L. Chan {et~al.}(2025)Chan, Seo, Li, Fong, \& Ezquiaga}]{Chan:2024qmb}
Chan, J. C.~L., Seo, E., Li, A. K.~Y., Fong, H., \& Ezquiaga, J.~M. 2025, \bibinfo{title}{{Detectability of lensed gravitational waves in matched-filtering searches},} Phys. Rev. D, 111, 084019, \dodoi{10.1103/PhysRevD.111.084019}

\bibitem[{A. Chen {et~al.}(2024)Chen, Cremonese, Ezquiaga, \& Keitel}]{Chen:2024xal}
Chen, A., Cremonese, P., Ezquiaga, J.~M., \& Keitel, D. 2024, \bibinfo{title}{{Invariance transformations in wave-optics lensing: implications for gravitational-wave astrophysics and cosmology},} \doarXiv{2408.03856}

\bibitem[{M.~H.-Y. Cheung {et~al.}(2024)Cheung, Ng, Zumalac{\'a}rregui, \& Berti}]{Cheung:2024ugg}
Cheung, M. H.-Y., Ng, K. K.~Y., Zumalac{\'a}rregui, M., \& Berti, E. 2024, \bibinfo{title}{{Probing minihalo lenses with diffracted gravitational waves},} Phys. Rev. D, 109, 124020, \dodoi{10.1103/PhysRevD.109.124020}

\bibitem[{M.~H.-Y. Cheung {et~al.}(2026)Cheung, Wadekar, Zaldarriaga, Venumadhav, Roulet, \& Mehta}]{Cheung:2026pky}
Cheung, M. H.-Y., Wadekar, D., Zaldarriaga, M., {et~al.} 2026, \bibinfo{title}{{The diffraction-lensing interpretation of GW231123 with astrophysical priors},} \doarXiv{2607.21834}

\bibitem[{Q. Chu {et~al.}(2022)Chu {et~al.}}]{Chu:2020pjv}
Chu, Q., {et~al.} 2022, \bibinfo{title}{{SPIIR online coherent pipeline to search for gravitational waves from compact binary coalescences},} Phys. Rev. D, 105, 024023, \dodoi{10.1103/PhysRevD.105.024023}

\bibitem[{L.~S. Collaboration {et~al.}(2025)Collaboration, Collaboration, \& Collaboration}]{LVK:2025a}
Collaboration, L.~S., Collaboration, V., \& Collaboration, K. 2025, GWTC-4.0: Candidate data release, Zenodo, \dodoi{10.5281/zenodo.16053641}

\bibitem[{P. Cremonese {et~al.}(2021)Cremonese, Ezquiaga, \& Salzano}]{Cremonese:2021puh}
Cremonese, P., Ezquiaga, J.~M., \& Salzano, V. 2021, \bibinfo{title}{{Breaking the mass-sheet degeneracy with gravitational wave interference in lensed events},} Phys. Rev. D, 104, 023503, \dodoi{10.1103/PhysRevD.104.023503}

\bibitem[{L. {Dai} {\&} T. {Venumadhav}(2017){Dai} \& {Venumadhav}}]{Dai:2017huk}
{Dai}, L., \& {Venumadhav}, T. 2017, \bibinfo{title}{{On the waveforms of gravitationally lensed gravitational waves},} arXiv e-prints, arXiv:1702.04724, \dodoi{10.48550/arXiv.1702.04724}

\bibitem[{T. Dal~Canton {et~al.}(2021)Dal~Canton, Nitz, Gadre, Cabourn~Davies, Villa-Ortega, Dent, Harry, \& Xiao}]{DalCanton:2020vpm}
Dal~Canton, T., Nitz, A.~H., Gadre, B., {et~al.} 2021, \bibinfo{title}{{Real-time Search for Compact Binary Mergers in Advanced LIGO and Virgo's Third Observing Run Using PyCBC Live},} Astrophys. J., 923, 254, \dodoi{10.3847/1538-4357/ac2f9a}

\bibitem[{J.~M. Ezquiaga {et~al.}(2021)Ezquiaga, Holz, Hu, Lagos, \& Wald}]{Ezquiaga:2020gdt}
Ezquiaga, J.~M., Holz, D.~E., Hu, W., Lagos, M., \& Wald, R.~M. 2021, \bibinfo{title}{{Phase effects from strong gravitational lensing of gravitational waves},} Phys. Rev. D, 103, 6, \dodoi{10.1103/PhysRevD.103.064047}

\bibitem[{J.~M. Ezquiaga {et~al.}(2025)Ezquiaga, Lo, \& Vujeva}]{Ezquiaga:2025gkd}
Ezquiaga, J.~M., Lo, R. K.~L., \& Vujeva, L. 2025, \bibinfo{title}{{Diffraction around caustics in gravitational wave lensing},} Phys. Rev. D, 112, 043544, \dodoi{10.1103/tkk2-x9st}

\bibitem[{M. Fairbairn {et~al.}(2023)Fairbairn, Urrutia, \& Vaskonen}]{Fairbairn:2022xln}
Fairbairn, M., Urrutia, J., \& Vaskonen, V. 2023, \bibinfo{title}{Microlensing of gravitational waves by dark matter structures,} Journal of Cosmology and Astroparticle Physics, 2023, 007

\bibitem[{A. Garoffolo {\&} G. Tasinato(2026)Garoffolo \& Tasinato}]{Garoffolo:2026hyx}
Garoffolo, A., \& Tasinato, G. 2026, \bibinfo{title}{{Wave-optics gravitational wave lensing in modified gravity},} Phys. Rev. D, 114, 064059, \dodoi{10.1103/smzt-5v1s}

\bibitem[{S. Gholap {et~al.}(2026)Gholap, Dhurandhar, \& Kapadia}]{Gholap:2026fjy}
Gholap, S., Dhurandhar, S.~V., \& Kapadia, S.~J. 2026, \bibinfo{title}{{A product template bank to search for compact binary coalescences microlensed by isolated point mass lenses},} \doarXiv{2607.20661}

\bibitem[{S. Goyal {et~al.}(2025)Goyal, Villarrubia-Rojo, \& Zumalacarregui}]{Goyal:2025eqo}
Goyal, S., Villarrubia-Rojo, H., \& Zumalacarregui, M. 2025, \bibinfo{title}{{Across the Universe: GW231123 as a magnified and diffracted black hole merger},} \doarXiv{2512.17631}

\bibitem[{O.~A. Hannuksela {et~al.}(2019)Hannuksela, Haris, Ng, Kumar, Mehta, Keitel, Li, \& Ajith}]{Hannuksela:2019kle}
Hannuksela, O.~A., Haris, K., Ng, K. K.~Y., {et~al.} 2019, \bibinfo{title}{{Search for gravitational lensing signatures in LIGO-Virgo binary black hole events},} Astrophys. J. Lett., 874, L2, \dodoi{10.3847/2041-8213/ab0c0f}

\bibitem[{S. Harikumar {et~al.}(2026)Harikumar, Askar, Bejger, Biesiada, Hendry, \& Janquart}]{Harikumar:2026pnu}
Harikumar, S., Askar, A., Bejger, M., {et~al.} 2026, \bibinfo{title}{{Probing globular clusters parameters through gravitational wave lensing with stellar-mass black hole binaries},} \doarXiv{2606.31531}

\bibitem[{Q. Hu {et~al.}(2025)Hu, Narola, Heynen, Wright, Veitch, Janquart, \& Van Den~Broeck}]{Hu:2025lhv}
Hu, Q., Narola, H., Heynen, J., {et~al.} 2025, \bibinfo{title}{{GW231123: Overlapping Gravitational Wave Signals?},} \doarXiv{2512.17550}

\bibitem[{Y.-J. Huang {et~al.}(2025)Huang {et~al.}}]{Huang:2025pkd}
Huang, Y.-J., {et~al.} 2025, \bibinfo{title}{{SGNL: Scalable Low-Latency Gravitational Wave Detection Pipeline for Compact Binary Mergers},} \doarXiv{2511.04730}

\bibitem[{J. Janquart {et~al.}(2023)Janquart {et~al.}}]{Janquart:2023mvf}
Janquart, J., {et~al.} 2023, \bibinfo{title}{{Follow-up analyses to the O3 LIGO{\textendash}Virgo{\textendash}KAGRA lensing searches},} Mon. Not. Roy. Astron. Soc., 526, 3832, \dodoi{10.1093/mnras/stad2909}

\bibitem[{J. Janquart {et~al.}(2024)Janquart {et~al.}}]{Janquart:2024ztv}
Janquart, J., {et~al.} 2024, \bibinfo{title}{{What is the nature of GW230529? An exploration of the gravitational lensing hypothesis},} \dodoi{10.1093/mnras/staf049}

\bibitem[{S. Jung {\&} C.~S. Shin(2019)Jung \& Shin}]{Jung:2017flg}
Jung, S., \& Shin, C.~S. 2019, \bibinfo{title}{{Gravitational-Wave Fringes at LIGO: Detecting Compact Dark Matter by Gravitational Lensing},} Phys. Rev. Lett., 122, 041103, \dodoi{10.1103/PhysRevLett.122.041103}

\bibitem[{S. Khan {et~al.}(2016)Khan, Husa, Hannam, Ohme, P{\"u}rrer, Jim{\'e}nez~Forteza, \& Boh{\'e}}]{Khan:2015jqa}
Khan, S., Husa, S., Hannam, M., {et~al.} 2016, \bibinfo{title}{{Frequency-domain gravitational waves from nonprecessing black-hole binaries. II. A phenomenological model for the advanced detector era},} Phys. Rev. D, 93, 044007, \dodoi{10.1103/PhysRevD.93.044007}

\bibitem[{{}{LIGO Scientific Collaboration} {et~al.}(2018){LIGO Scientific Collaboration}, {Virgo Collaboration}, \& {KAGRA Collaboration}}]{lalsuite}
{LIGO Scientific Collaboration}, {Virgo Collaboration}, \& {KAGRA Collaboration}. 2018, {LVK} {A}lgorithm {L}ibrary - {LALS}uite,, Free software (GPL) \dodoi{10.7935/GT1W-FZ16}

\bibitem[{{}{LIGO Scientific, Virgo, and KAGRA Collaboration}(2025){LIGO Scientific, Virgo, and KAGRA Collaboration}}]{LVK:2025c}
{LIGO Scientific, Virgo, and KAGRA Collaboration}. 2025, {GW231123: a Binary Black Hole Merger with Total Mass 190-265\,$M_\odot$ --- Data Release}, Zenodo, \dodoi{10.5281/zenodo.15832843}

\bibitem[{A. Liu {et~al.}(2024)Liu, Chandramouli, Hannuksela, Yunes, \& Li}]{Liu:2024xxn}
Liu, A., Chandramouli, R.~S., Hannuksela, O.~A., Yunes, N., \& Li, T. G.~F. 2024, \bibinfo{title}{{Millilensing induced systematic biases in parameterized tests of General Relativity},} \doarXiv{2410.21738}

\bibitem[{A. Liu {\&} K. Kim(2024)Liu \& Kim}]{Liu:2023emk}
Liu, A., \& Kim, K. 2024, \bibinfo{title}{{Can we discern millilensed gravitational-wave signals from signals produced by precessing binary black holes with ground-based detectors?},} Phys. Rev. D, 110, 123008, \dodoi{10.1103/PhysRevD.110.123008}

\bibitem[{A. Liu {et~al.}(2023)Liu, Wong, Leong, More, Hannuksela, \& Li}]{Liu:2023ikc}
Liu, A., Wong, I. C.~F., Leong, S. H.~W., {et~al.} 2023, \bibinfo{title}{{Exploring the hidden Universe: a novel phenomenological approach for recovering arbitrary gravitational-wave millilensing configurations},} Mon. Not. Roy. Astron. Soc., 525, 4149, \dodoi{10.1093/mnras/stad1302}

\bibitem[{R.~K.~L. Lo {et~al.}(2025)Lo, Vujeva, Ezquiaga, \& Chan}]{Lo:2024wqm}
Lo, R. K.~L., Vujeva, L., Ezquiaga, J.~M., \& Chan, J. C.~L. 2025, \bibinfo{title}{{Observational Signatures of Highly Magnified Gravitational Waves from Compact Binary Coalescence},} Phys. Rev. Lett., 134, 151401, \dodoi{10.1103/PhysRevLett.134.151401}

\bibitem[{R. Magee {et~al.}(2024)Magee, Isi, Payne, Chatziioannou, Farr, Pratten, \& Vitale}]{Magee:2023muf}
Magee, R., Isi, M., Payne, E., {et~al.} 2024, \bibinfo{title}{{Impact of selection biases on tests of general relativity with gravitational-wave inspirals},} Phys. Rev. D, 109, 023014, \dodoi{10.1103/PhysRevD.109.023014}

\bibitem[{N. Menadeo {\&} M. Zumalac{\'a}rregui(2025)Menadeo \& Zumalac{\'a}rregui}]{Menadeo:2024uoq}
Menadeo, N., \& Zumalac{\'a}rregui, M. 2025, \bibinfo{title}{{Gravitational wave propagation beyond general relativity: Geometric optic expansion and lens-induced dispersion},} Phys. Rev. D, 111, 104022, \dodoi{10.1103/PhysRevD.111.104022}

\bibitem[{C. Messick {et~al.}(2017)Messick {et~al.}}]{Messick:2017a}
Messick, C., {et~al.} 2017, \bibinfo{title}{{Analysis Framework for the Prompt Discovery of Compact Binary Mergers in Gravitational-wave Data},} Phys. Rev. D, 95, 042001, \dodoi{10.1103/PhysRevD.95.042001}

\bibitem[{A. Mishra {et~al.}(2024)Mishra, Meena, More, \& Bose}]{Mishra:2023ddt}
Mishra, A., Meena, A.~K., More, A., \& Bose, S. 2024, \bibinfo{title}{{Exploring the impact of microlensing on gravitational wave signals: Biases, population characteristics, and prospects for detection},} Mon. Not. Roy. Astron. Soc., 531, 764, \dodoi{10.1093/mnras/stae836}

\bibitem[{T.~T. Nakamura(1998)Nakamura}]{Nakamura:1997sw}
Nakamura, T.~T. 1998, \bibinfo{title}{{Gravitational lensing of gravitational waves from inspiraling binaries by a point mass lens},} Phys. Rev. Lett., 80, 1138, \dodoi{10.1103/PhysRevLett.80.1138}

\bibitem[{E. Pizzati {et~al.}(2022)Pizzati, Sachdev, Gupta, \& Sathyaprakash}]{Pizzati:2021apa}
Pizzati, E., Sachdev, S., Gupta, A., \& Sathyaprakash, B. 2022, \bibinfo{title}{{Toward inference of overlapping gravitational-wave signals},} Phys. Rev. D, 105, 104016, \dodoi{10.1103/PhysRevD.105.104016}

\bibitem[{S. Savastano {et~al.}(2023)Savastano, Tambalo, Villarrubia-Rojo, \& Zumalacarregui}]{Savastano:2023spl}
Savastano, S., Tambalo, G., Villarrubia-Rojo, H., \& Zumalacarregui, M. 2023, \bibinfo{title}{{Weakly lensed gravitational waves: Probing cosmic structures with wave-optics features},} Phys. Rev. D, 108, 103532, \dodoi{10.1103/PhysRevD.108.103532}

\bibitem[{X. Shan {et~al.}(2026)Shan, Wang, Mao, Yang, \& Zumalac{\'a}rregui}]{Shan:2026beq}
Shan, X., Wang, Y.-F., Mao, S., Yang, H., \& Zumalac{\'a}rregui, M. 2026, \bibinfo{title}{{Gravitational lensing of gravitational waves by galaxy clusters},} \doarXiv{2609.16891}

\bibitem[{R. Takahashi {\&} T. Nakamura(2003)Takahashi \& Nakamura}]{Takahashi:2003ix}
Takahashi, R., \& Nakamura, T. 2003, \bibinfo{title}{{Wave effects in gravitational lensing of gravitational waves from chirping binaries},} Astrophys. J., 595, 1039, \dodoi{10.1086/377430}

\bibitem[{G. Tambalo {et~al.}(2023)Tambalo, Zumalac\'arregui, Dai, \& Cheung}]{Tambalo:2022wlm}
Tambalo, G., Zumalac\'arregui, M., Dai, L., \& Cheung, M. H.-Y. 2023, \bibinfo{title}{{Gravitational wave lensing as a probe of halo properties and dark matter},} Phys. Rev. D, 108, 103529, \dodoi{10.1103/PhysRevD.108.103529}

\bibitem[{H. Ubach {et~al.}(2025)Ubach, Gieles, \& Miralda-Escud{\'e}}]{Ubach:2025dob}
Ubach, H., Gieles, M., \& Miralda-Escud{\'e}, J. 2025, \bibinfo{title}{{Constraining the environment of compact binary mergers with self-lensing signatures},} Phys. Rev. D, 112, 083026, \dodoi{10.1103/ql7q-t6wc}

\bibitem[{S.~A. Usman {et~al.}(2016)Usman {et~al.}}]{Usman:2015kfa}
Usman, S.~A., {et~al.} 2016, \bibinfo{title}{{The PyCBC search for gravitational waves from compact binary coalescence},} Class. Quant. Grav., 33, 215004, \dodoi{10.1088/0264-9381/33/21/215004}

\bibitem[{V. Varma {et~al.}(2019)Varma, Field, Scheel, Blackman, Gerosa, Stein, Kidder, \& Pfeiffer}]{Varma:2019csw}
Varma, V., Field, S.~E., Scheel, M.~A., {et~al.} 2019, \bibinfo{title}{{Surrogate models for precessing binary black hole simulations with unequal masses},} Phys. Rev. Research., 1, 033015, \dodoi{10.1103/PhysRevResearch.1.033015}

\bibitem[{T. Venumadhav {et~al.}(2019)Venumadhav, Zackay, Roulet, Dai, \& Zaldarriaga}]{Venumadhav:2019tad}
Venumadhav, T., Zackay, B., Roulet, J., Dai, L., \& Zaldarriaga, M. 2019, \bibinfo{title}{{New search pipeline for compact binary mergers: Results for binary black holes in the first observing run of Advanced LIGO},} Phys. Rev. D, 100, 023011, \dodoi{10.1103/PhysRevD.100.023011}

\bibitem[{L. Vujeva {et~al.}(2026)Vujeva, Ezquiaga, Lo, \& Zwick}]{Vujeva:2026xfh}
Vujeva, L., Ezquiaga, J.~M., Lo, R. K.~L., \& Zwick, L. 2026, \bibinfo{title}{{Probability of gravitational-wave lensing by intermediate-mass black holes and globular clusters},} \doarXiv{2608.07667}

\bibitem[{B. Wang {\&} T. Yang(2026)Wang \& Yang}]{Wang:2026yjk}
Wang, B., \& Yang, T. 2026, \bibinfo{title}{{GW231123: False Massive Graviton Signatures from Unmodeled Point-Mass Lensing},} \doarXiv{2604.08179}

\bibitem[{Y. Wang {et~al.}(1996{\natexlab{a}})Wang, Stebbins, \& Turner}]{Wang:1996as}
Wang, Y., Stebbins, A., \& Turner, E.~L. 1996{\natexlab{a}}, \bibinfo{title}{{Gravitational lensing of gravitational waves from merging neutron star binaries},} Phys. Rev. Lett., 77, 2875, \dodoi{10.1103/PhysRevLett.77.2875}

\bibitem[{K. Wette(2020)Wette}]{swiglal}
Wette, K. 2020, \bibinfo{title}{{SWIGLAL: Python and Octave interfaces to the LALSuite gravitational-wave data analysis libraries},} SoftwareX, 12, 100634, \dodoi{10.1016/j.softx.2020.100634}

\bibitem[{M. Zumalac{\'a}rregui {\&} X. Shan(2026)Zumalac{\'a}rregui \& Shan}]{Zumalacarregui:2026uqs}
Zumalac{\'a}rregui, M., \& Shan, X. 2026, \bibinfo{title}{{Effective description of lensed gravitational waves diffracted by stellar fields},} Phys. Rev. D, 114, 063038, \dodoi{10.1103/d5fx-xgd2}

\end{thebibliography}
\end{document}